*Review Article*

# Physical Environment of Accreting Neutron Stars

**J. Wang**

*Institute of Astronomy and Space Science, Sun Yat-Sen University, Guangzhou 510275, China*

Correspondence should be addressed to J. Wang; wangjing6@mail.sysu.edu.cn

Received 8 November 2015; Revised 20 February 2016; Accepted 3 March 2016

Academic Editor: Elmetwally Elabbasy

Neutron stars (NSs) powered by accretion, which are known as accretion-powered NSs, always are located in binary systems and manifest themselves as X-ray sources. Physical processes taking place during the accretion of material from their companions form a challenging and appealing topic, because of the strong magnetic field of NSs. In this paper, we review the physical process of accretion onto magnetized NS in X-ray binary systems. We, firstly, give an introduction to accretion-powered NSs and review the accretion mechanism in X-ray binaries. This review is mostly focused on accretion-induced evolution of NSs, which includes scenario of NSs both in high-mass binaries and in low-mass systems.

## 1. Introduction to Accretion-Powered Neutron Star

Since the discovery for the first extrasolar X-ray source, Scorpius X-1 in 1962 [1], a new field of astronomy—accreting compact objects in the galaxy—has arisen, which offers unique insight into the physics at extreme conditions. So far, more and more bright galactic X-ray sources have been discovered, which show a clear concentration towards both galactic center and galactic plane. They are highly variable on timescales of seconds or less and display typical source luminosity of $10^{34}$–$10^{38}$ ergs s$^{-1}$, in the energy band of 1–10 keV. It was suggested that the strong galactic X-ray sources are NSs accreting material from their companions in close binary systems [2]. The detection of coherent and regular pulsations from the accreting X-ray sources Centaurus X-3 [3, 4] and Hercules X-1 [5] by Uhuru in 1971 provided the first evidence that the compact objects in many of these sources are indeed accreting neutron stars. Since then, accretion onto rotating NSs have been nicely confirmed as the standard picture of strongest galactic X-ray sources, and the X-ray pulsation periods are the spin periods of X-ray pulsars.

Accretion remains the only viable source of power to the binary X-ray pulsars as a whole [6–8]. For a NS of mass $M \sim M_\odot$, and of radius $R = 10^6$ cm, the quantity of energy released by matter of mass $m$ falling into its deep gravitational potential well amounts to

$$\Delta E_{\mathrm{acc}} = \frac{GMm}{R}, \quad (1)$$

where $G$ is the gravitational constant. This is up to $\sim 10^{20}$ ergs g$^{-1}$ of accreted matter, that is, about a tenth of its rest-mass energy ($\sim 0.1c^2$, with the speed of light $c$), which makes accretion as an ideal source of power. Since each unit of accreted mass releases an amount of gravitational potential energy $GM/R$ when it reaches the NS surface, the luminosity generated by the accretion process is given by

$$L_{\mathrm{acc}} = \frac{GM\dot{m}}{R}, \quad (2)$$

where $\dot{m}$ is the accretion rate. Accordingly, to generate a typical luminosity of about $10^{37}$ ergs s$^{-1}$, it requires an accretion rate of $\sim 10^{17}$ g s$^{-1}$ $\sim 10^{-9} M_\odot$ yr$^{-1}$.

During the accretion process, radiation passes through accretion flow and influences its dynamics in the case of a sufficiently large luminosity. When the outward pressure of radiation exceeds the inward gravitational attraction, the infalling flow will be halted, which implies a critical luminosity, Eddington luminosity [9]:

$$L_{\mathrm{Edd}} \approx 1.3 \times 10^{38} \frac{M}{M_\odot} \text{ ergs s}^{-1}, \quad (3)$$

which, for spherically symmetric accretion, corresponds to a limit on a steady accretion rate, the Eddington accretion rate:

$$\dot{M}_{\rm Edd} \approx 10^{18} R_6\,{\rm g\,s^{-1}} \approx 1.5 \times 10^{-8} M_{\odot} R_6\,{\rm yr^{-1}}. \quad (4)$$

Lower and higher accretion rates than this critical value are called the "sub-" and "super-"Eddington accretion, respectively.

Phenomenologically, over 95% of the NS X-ray binaries fall into two distinct categories, that is, low-mass X-ray binary (LMXB) and high-mass X-ray binary (HMXB) [10, 11]. Wind-fed system, in which the companion of NS loses mass in the form of a stellar wind, requires a massive companion with mass $\geq 10 M_{\odot}$ to drive the intense stellar wind and thus appears to be high-mass system, such as Centaurus X-3. NSs in HMXBs display relatively hard X-ray spectra with a peak energy larger than 15 keV [12] and tend to manifest as regular X-ray pulsars, with spin periods of $1$–$10^3$ s and magnetic fields of $10^{11}$–$10^{13}$ G [13]. The companions in HMXBs are bright and luminous early-type Be or OB supergiant stars, which have masses larger than 10 solar masses and are short lived and belong to the youngest stellar population in the galaxy, with ages of $\sim 10^5$–$10^7$ yrs [14–16]. They are distributed close to the galactic plane. On the other hand, X-ray binaries, with companion mass, $\leq M_{\odot}$, are categorized as low-mass X-ray binaries (LMXBs), in which mass transfer takes place by Roche-lone overflow. This transfer is driven either by losing angular momentum due to gravitational radiation (for systems of very small masses and orbital separations) and magnetic braking (for systems of orbital periods $P_{\rm orb} \leq$ 2 days) or by the evolution of the companion star (for systems of $P_{\rm orb} \geq$ 2 days). Most NSs in LMXBs are X-ray busters [17], with relatively soft X-ray spectra of <10 keV in exponential fittings [11]. In LMXBs, the NSs possess relatively weak surface magnetic fields of $10^{8\text{-}9}$ G and short spin periods of a few milliseconds [18], while the companions of NSs are late-type main-sequence stars, white dwarfs, or subgiant stars with F-G type spectra [19, 20]. In space the LMXBs are concentrated towards the galactic center and have a fairly wide distribution around the galactic plane, which characterize a relatively old population, with ages of $(0.5$–$1.5) \times 10^{10}$ yrs. A few of strong galactic X-ray sources, for example, Hercules X-1, are assigned to be intermediate-mass X-ray binaries [21, 22], in which the companions of NSs have masses of 1-2$M_{\odot}$. Intermediate-mass systems are rare, since mass transfer via Roche lobe is unstable and would lead to a very quick ($\sim 10^3$–$10^5$ yrs) evolution of the system, while, in the case of stellar wind accretion, the mass accretion rate is very low and the system would be very dim and hardly detectable [23].

## 2. Physical Processes in Accreting Neutron Stars

*2.1. Mass Transfer in Binary Systems.* The mass transfer from the companion to NS is either due to a stellar wind or to Roche lobe overflow (RLOF), as described above [7, 24].

A massive companion star, at some evolutionary phase, ejects much of its mass in the form of intense stellar wind driven by radiation [25, 26]. The wind material leaves surface uniformly in all directions. Some of material will be captured gravitationally by the NS. This mechanism is called stellar wind accretion, and the corresponding systems are wind-fed systems [7].

In the course of its evolution, the optical companion star may increase in radius to a point where the gravitational pull of the companion can remove the outer layers of its envelope, which is the RLOF phase [9, 27]. In addition to the swelling of the companion, this situation can also come about, due to, for example, the decrease of binary separation as a consequence of magnetically coupled wind mass loss or gravitational radiation. Some of material begins to flow over to the Roche lobe of NS through the inner Lagrangian point, without losing energy. Because of the large angular momentum, the matter enters an orbit at some distance from the NS. Subsequent batches of outflowing matter from companion have nearly the same initial conditions and hence fall on the same orbit. As a result, a ring of increasing density is formed by the gas near the NS. Because of the mutual collisions of individual density or turbulence cells, the angular momentum in the ring is redistributed. The ring spreads out into a disk in which the matter rotates differentially. Gradually, the motion turns out into steady-state disk accretion on a time scale determined by the viscosity. Some optical observations suggested that mass transfer by ROLF is taking place or perhaps mixed with that by stellar wind accretion in some systems.

*2.2. Accretion Regimes.* Because of the different angular momentum carried by the accreted material and the motion of the NS relative to the sound speed $c_s$ in the medium, for accretion onto a NS, several modes are possible [28].

*2.2.1. Spherically Symmetric Accretion [29].* The NS hardly moves relative to the medium in its vicinity; that is, $v_{\infty} \ll c_s$. The gas, of uniform density and pressure, is at rest and does not possess significant angular momentum at infinity and falls freely, with spherically symmetrical and steady motion, towards the NS.

*2.2.2. Cylindrical Accretion [30].* Although the angular momentum is small, the NS velocity is comparable with, or larger than, the sound speed in the accreted matter; that is, $v_{\infty} \gtrsim c_s$. In this case, a conical shock form will present as the result of pressure effects.

*2.2.3. Disk Accretion [6, 31].* In a binary system, if the material supplied by the companion of the NS has a large angular momentum due to the orbital motion, we must take both the gravitational force and centrifugal forces into consideration. As a result, an accretion disk ultimately forms around a NS in the binary plane.

In a number of astrophysical cases [32], the motion of accreted material in the vicinity of a compact object presents two-stream accretion, that is, a quasi-spherically symmetric flux of matter in addition to the accretion disk [33].

*2.3. Accretion from a Stellar Wind.* Mass loss in the case of OB supergiant stars is driven by radiation pressure [34]. The wind material is accelerated outwards from stellar surface to a final velocity $v_\infty$, which is related to the escape velocity $v_{esc} = \sqrt{2GM_{OB}/R_{OB}}$, for an OB companion with mass $M_{OB}$ and radius $R_{OB}$. In most cases, $v_\infty \approx 3v_{esc}$. In these circumstances, the stellar wind is intense, characterized by a mass loss rate $\dot{M}_w \sim 10^{-6}$–$10^{-5} M_\odot\ \mathrm{yr}^{-1}$, and highly supersonic, with a wind velocity of $v_w \sim 1000$–$2000\ \mathrm{km\ s^{-1}}$, which is much larger than the orbital velocity $v_{orb} \sim 200\ \mathrm{km\ s^{-1}}$ at $2R_{OB}$.

*2.3.1. Characteristic Radii.* In the wind accretion theory, three characteristic radii [7, 8, 35] can be defined as follows.

*Accretion Radius.* Only part of stellar wind within a certain radius will be captured by the gravitational field of the NS, whereas the wind material outside that radius will escape. This radius, called the accretion radius $r_{acc}$ [28], can be defined by noting that material will only be accreted if it has a kinetic energy lower than the potential energy in the gravitational potential well of NS; that is,

$$r_{acc} = \frac{2GM}{v_w^2} \sim 10^{10} \frac{M}{M_\odot} v_{w,8}^{-2}\ \mathrm{cm}, \tag{5}$$

where $v_{w,8}$ is a wind velocity in units of $1000\ \mathrm{km\ s^{-1}}$ ($v_{w,8} = v_w/(10^8\ \mathrm{cm\ s^{-1}})$). We consider now two cases: $r_{mag}$ smaller than $r_{acc}$ and $r_{mag}$ larger than $r_{acc}$.

*Magnetospheric Radius.* (i) $r_{mag} < r_{acc}$: when accreting, the electromagnetic field of NS will hinder the plasma from falling all the way to its surface. The inflow will come to stop at certain distance, at which the magnetic pressure of NS can balance the ram pressure of accreting flow. Here the magnetosphere forms at a magnetospheric radius $r_{mag}$ [36, 37], which is defined by

$$\frac{B\left(r_{mag}\right)^2}{8\pi} = \rho\left(r_{mag}\right) v\left(r_{mag}\right)^2, \tag{6}$$

where $B(r_{mag}) \sim \mu/r_{mag}^3$ is the magnetic field strength at $r_{mag}$ and $\rho(r_{mag})$ and $v(r_{mag})$ are wind density and wind velocity at $r_{mag}$, respectively. $\mu$ is the magnetic moment of the NS. (ii) $r_{mag} > r_{acc}$: here the inflows cannot experience a significant gravitational field. Assuming a nonmagnetized spherically symmetric wind [36], the wind density near the NS $\rho(r_{mag})$ given by $\rho(r_{mag}) \approx \dot{M}_w/4\pi a^2 v_w$ [37, 38], where $a$ is the orbital separation between two components and $a \gg r_{mag}$, is assumed. Therefore, the magnetospheric radius in this case is given by setting $\rho(r_{mag})v_w^2 = B(r_{mag})^2/8\pi$, which yields

$$r_{mag} \sim 10^{10} \dot{M}_{w,-6}^{-1/6} v_{w,8}^{-1/6} a_{10}^{1/3} \mu_{33}^{1/3}\ \mathrm{cm}, \tag{7}$$

where $\dot{M}_{w,-6} = \dot{M}_w/(10^{-6}M_\odot/\mathrm{yr})$, $\mu_{33} = \mu/(10^{33}\ \mathrm{G\ cm^3})$, and $a_{10}$ is related to the orbital separation as $a \sim 10^{12} a_{10}$ cm $\sim 10^{12} P_{b,10}^{2/3} M_{30}^{1/3}$ cm. $P_{b,10}$ is the orbital period $P_b$ in units of 10 days, and $M_{30}$ is the companion mass in units of 30 solar masses. When $r_{mag} < r_{acc}$, the gravitational field of NS dominates the falling of wind matter. The wind density and wind velocity at $r_{mag}$ can therefore be taken as $\rho(r_{mag}) = \dot{M}/v(r_{mag})4\pi r_{mag}^2$ and $v(r_{mag}) = \sqrt{2GM/r_{mag}}$, respectively. The accretion rate $\dot{M}$ depends on $r_{acc}$ and $a$, according to [9]

$$\frac{\dot{M}}{\dot{M}_w} \sim \frac{r_{acc}^2}{(4a^2)} \sim 10^{-5} \left(\frac{M}{M_\odot}\right)^2 v_{w,8}^{-4} a_{10}^{-2}. \tag{8}$$

Then the magnetospheric radius is given by using (6) which yields

$$r_{mag} \sim 10^{10} \left(\frac{M}{M_\odot}\right)^{-5/7} \dot{M}_{w,-6}^{-2/7} v_{w,8}^{8/7} a_{10}^{4/7} \mu_{33}^{4/7}\ \mathrm{cm}. \tag{9}$$

*Corotation Radius.* A corotation radius $r_{cor}$ can be defined when the spin angular velocity ($\Omega_s = 2\pi/P_s$) of NS is equal to the Keplerian angular velocity ($\Omega_k = \sqrt{GM/r^3}$) of matter being accreted:

$$r_{cor} \sim 10^{10} \left(\frac{M}{M_\odot}\right)^{1/3} P_{s,3}^{2/3}\ \mathrm{cm}, \tag{10}$$

where $P_{s,3}$ is the spin period of the neutron star $P_s$ in units of $10^3$ s.

*2.3.2. Accretion onto a Magnetized NS.* The stellar wind flows approximately spherically symmetric outside the accretion radius. When approaching the NS, according to the relative dimensions of three characteristic radii, different accretion regimes can be identified.

*(I) Magnetic Inhibition: Propeller.* In a system in which the magnetospheric radius is larger than the accretion radius ($r_{mag} > r_{acc}$), the stellar wind, flowing around the NS, may not experience a significant gravitational field and interacts directly with the magnetosphere. In this case, very little material penetrates the magnetospheric radius and will be accreted by NS [24]. Most of wind material is ejected by the rotation energy of the NS. This is called the propeller mechanism. The NS in this situation behaves like a propeller and spins down due to dissipation of rotational energy resulting from the interaction between the magnetosphere and stellar wind. The spin-down torque is expressed as [39, 40]

$$T_{m,pro} = -\frac{\kappa_t \mu^2}{r_{mag}^3}, \tag{11}$$

where $\kappa_t \leq 1$ is a dimensionless parameter of order unity [37]. The equation governing the spin evolution can be written as [7, 41]

$$2\pi I \frac{d}{dt}\frac{1}{P_s} = T, \tag{12}$$

where $I$ is the moment of inertia of NS and $T$ is the total torque imposed on NS. Therefore, the spin-down rate will be

$$\dot{P}_{s,m} \sim 10^{-6} \kappa_t \mu_{33} I_{45}^{-1} P_{s,3}^2 a_{10}^{-1} \dot{M}_{w,-6}^{1/2} v_{w,8}^{1/2}\ \mathrm{s\ s^{-1}}, \tag{13}$$

where $I_{45}$ is the moment of inertia in units of $10^{45}\,\mathrm{g\,cm^2}$. Accordingly, if the NS can spin down to a period of 1000 s under this torque, the spin-down timescale is given by

$$\tau_{m,\mathrm{pro}} = P_s/\dot{P}_{s,\mathrm{pur}} \sim 10^3 \kappa_t^{-1} \mu_{33}^{-1} I_{45} P_{s,3}^{-1} a_{10} \dot{M}_{w,-6}^{-1/2} v_{w,8}^{-1/2}\ \mathrm{yr}. \quad (14)$$

In this regime, the spin period $P_s$ must satisfy

$$P_s > 11.2 \frac{M}{M_\odot} v_{w,8}^{-2}\ \mathrm{s}. \quad (15)$$

In addition, the mechanism of magnetic inhibition of accretion only occurs before the centrifugal barrier acts; that is, the condition $r_{\mathrm{cor}} > r_{\mathrm{mag}}$ must be satisfied, which gives

$$P_s > 6.4 \times 10^3 \frac{M}{M_\odot} v_{w,8}^{-3}\ \mathrm{s}. \quad (16)$$

By imposing $r_{\mathrm{mag}} = r_{\mathrm{acc}}$, we get the minimum X-ray luminosity,

$$L_{x,m}(\mathrm{min}) \sim 10^{33} \left(\frac{M}{M_\odot}\right)^{-3} R_6^{-1} \mu_{33}^2 v_{w,8}^7\ \mathrm{ergs\,s^{-1}}, \quad (17)$$

and the corresponding maximum orbital period,

$$P_{\mathrm{orb},m}(\mathrm{max}) \sim 10^3 \left(\frac{M}{M_\odot}\right)^{9/2} M_{\mathrm{OB},30}^{-1/2} \mu_{33}^{-3/2} \dot{M}_{w,-6}^{3/4} v_{w,8}^{-33/4}\ \mathrm{days}. \quad (18)$$

For very fast rotating X-ray pulsars ($P_s \ll 1$ s), the magnetic pressure exerted by pulsar on wind material provides an additional mechanism for inhibiting accretion through $r_{\mathrm{acc}}$ [24]. The condition that magnetic pressure at $r_{\mathrm{acc}}$ counterbalances the wind ram pressure gives the minimum X-ray luminosity $L_{x,r}$(min) possible for accretion through $r_{\mathrm{acc}}$ to take place,

$$L_{x,r}(\mathrm{min}) \sim 10^{43} \mu_{33}^2 \left(\frac{P_s}{0.1\,\mathrm{s}}\right)^{-4} \frac{M}{M_\odot} R_6^{-1} v_{w,8}^{-1}\ \mathrm{ergs\,s^{-1}}. \quad (19)$$

Once the mass capture rate at $r_{\mathrm{acc}}$ is higher than that of (19), the flow will continue to be accreted by the NS until reaching the centrifugal barrier. This scenario is the magnetic inhibition of accretion like that of a radio pulsar [35].

*(II) Centrifugal Inhibition: Quasi-Spherical Capture.* If the magnetospheric radius is smaller than the accretion radius, the wind material penetrates through the accretion radius and halts at $r_{\mathrm{mag}}$. The captured material cannot penetrate any further because of a super-Keplerian drag exerted by the magnetic field of NS. The inflow in the region between $r_{\mathrm{acc}}$ and $r_{\mathrm{mag}}$ falls approximately in a spherical configuration. The NS behaves like a supersonic rotator, that is, the linear velocity of magnetosphere is much higher than the sound speed in wind clumps, which strongly shocks the flow at magnetospheric boundary and ejects some of material beyond the accretion radius, via propeller mechanism. This scenario corresponds to the propeller mechanism and imposes a spin-down torque on the NS, which is expressed by [32]

$$T_{c,\mathrm{pro}} = -\frac{\dot{M} v_w^2}{\Omega_s}. \quad (20)$$

In addition, due to the high speed of stellar wind, the accumulation rate of accreted matter is higher than the ejection rate. Consequently, more and more wind clumps are deposited outside the magnetosphere [42]. This regime is in connection with the spin-down evolution of young X-ray pulsars [24]. The minimum X-ray luminosity, at which accretion possibly occurs, is given by

$$L_{x,c}(\mathrm{min}) \sim 10^{43} R_6^{-1} \left(\frac{M}{M_\odot}\right)^{-2/3} \mu_{33}^2 \left(\frac{P_s}{1\,\mathrm{s}}\right)^{-7/3}\ \mathrm{ergs\,s^{-1}}. \quad (21)$$

An X-ray pulsar, with luminosity below this value, turns off as a result of the centrifugal mechanism. Correspondingly, a threshold in wind parameters is given by a relation between spin period and orbital period [35],

$$P_{s,c}(\mathrm{min}) \sim \left(\frac{M}{M_\odot}\right)^{-11/7} M_{30}^{2/7} \mu_{33}^{6/7} \left(\frac{P_{\mathrm{orb}}}{1\,\mathrm{day}}\right)^{4/7} \dot{M}_{w,-6}^{-3/7} v_{w,8}^{12/7}\ \mathrm{s}, \quad (22)$$

which is obtained by using the condition $r_{\mathrm{mag}} = r_{\mathrm{cor}}$, Kepler's third law, and a free-fall approximation for the wind material within $r_{\mathrm{acc}}$.

In the vicinity of the NS, the deposited wind matter is disordered because of the supersonic rotation, which causes turbulent motion around the magnetosphere. The velocity of the turbulent wind at the magnetospheric boundary is close to the sound speed [32]. Consequently, the accumulated matter can rotate either prograde or retrograde [43]. When flowing prograde, an accretion torque will transfer some angular momentum onto the NS and spin it up. The spin-up torque is

$$T_{c,\mathrm{acc},p} = \dot{M} \Omega_s r_{\mathrm{mag}}^2. \quad (23)$$

Therefore, the total torque is written as

$$T_{c,p} = T_{c,\mathrm{pro}} + T_{c,\mathrm{acc},p} = \dot{M} \Omega_s r_{\mathrm{mag}}^2 - \frac{\dot{M} v_w^2}{\Omega_s}. \quad (24)$$

The NS then behaves as a prograde propeller.

If the wind flows retrogradely, it imposes an inverted torque and spins down the NS, which is

$$T_{c,\mathrm{acc},r} = -T_{c,\mathrm{acc},p} = -\dot{M} \Omega_s r_{\mathrm{mag}}^2. \quad (25)$$

Accordingly, the total torque imposed on the NS in this scenario is

$$T_{c,r} = T_{c,\mathrm{pro}} + T_{c,\mathrm{acc},r} = -\frac{\dot{M} v_w^2}{\Omega_s} - \dot{M} \Omega_s r_{\mathrm{mag}}^2. \quad (26)$$

The NS is then a retrograde propeller and can spin down to a very long spin period [43]. Substituting (26) into (12), we can obtain the spin evolutionary law in the regime of retrograde propeller:

$$\dot{P}_{s,c,r} \sim 10^{-7} I_{45}^{-1} \left[ \dot{M}_{w,-6} v_{w,8}^{-2} a_{10}^{-2} P_{s,3}^{3} + \left( \frac{M}{M_\odot} \right)^{9/7} \cdot \dot{M}_{w,-6}^{5/7} a_{10}^{-10/7} v_{w,8}^{-20/7} \mu_{33}^{4/7} P_{s,3} \right] \mathrm{s\,s}^{-1}. \quad (27)$$

The NS can spin down to 1000 s during a timescale of

$$\tau_{c,r} \sim 10^{4} I_{45} \left[ \dot{M}_{w,-6} v_{w,8}^{-2} a_{10}^{-2} P_{s,3}^{2} + \left( \frac{M}{M_\odot} \right)^{9/7} \dot{M}_{w,-6}^{5/7} a_{10}^{-10/7} v_{w,8}^{-20/7} \mu_{33}^{4/7} \right]^{-1} \mathrm{yr}. \quad (28)$$

Therefore, the retrograde propeller can be an alternative mechanism for NSs with very long spin periods, such as supergiant fast X-ray transients [43].

If the stellar wind is not strong enough, the accreting matter is approximately in radial free fall as it approaches the magnetospheric boundary. The captured material cannot accumulate near the magnetosphere due to the supersonic rotation of NS. Consequently, the accretion torque contains only the spin-down torque imposed by the propeller mechanism, and the details are discussed by Urpin et al. [44, 45].

In the retrograde propeller phase, both interaction with magnetic fields by the propeller mechanism and inverted accretion contribute to the luminosity, which results in

$$L_{c,r} = \dot{M} v_w^2 + \frac{GM\dot{M}}{r_{\mathrm{mag}}} \sim 10^{32} \left[ 10 \left( \frac{M}{M_\odot} \right)^{2} \cdot \dot{M}_{w,-6} v_{w,8}^{-2} a_{10}^{-2} + \left( \frac{M}{M_\odot} \right)^{26/7} \cdot \dot{M}_{w,-6}^{9/7} v_{w,8}^{-36/7} a_{10}^{-18/7} \mu_{33}^{-4/7} \right] \mathrm{ergs\,s}^{-1}. \quad (29)$$

In addition, energy is also released through the shock formed near $r_{\mathrm{mag}}$, and the corresponding luminosity is discussed by Bozzo et al. [38].

*(III) Direct Wind Accretion: Accretor*. When $r_{\mathrm{acc}} > r_{\mathrm{mag}}$ and $r_{\mathrm{cor}} > r_{\mathrm{mag}}$, the captured material flows from the accretion radius and falls down directly toward the magnetosphere, where it is stopped by a collisionless shock. Because of a sub-Keplerian rotation at the magnetospheric boundary, the rotating NS cannot eject the wind material in this case, and the inflows experience a significant gravitational field and accumulate near the magnetosphere. Some of them penetrate the NS magnetosphere, presumably in a sporadic fashion, by means of Rayleigh-Taylor and Kelvin-Helmholtz instabilities [46]. Along the open field lines above the magnetic poles matter can also flow in relatively freely. Most accumulated matter forms an extended adiabatic, tenuous, and rotating atmosphere around the magnetosphere. The time-averaged X-ray luminosity is governed by the wind parameters at $r_{\mathrm{acc}}$,

$$L_x \sim 10^{33} R_6^{-1} \left( \frac{M}{M_\odot} \right)^{3} M_{30}^{-2/3} \left( \frac{P_{\mathrm{orb}}}{1\ \mathrm{day}} \right)^{-4/3} \cdot \dot{M}_{w,-6} v_{w,8}^{-4}\ \mathrm{ergs\,s}^{-1}, \quad (30)$$

assuming all gravitational potential energy is converted into X-ray luminosity.

When a gradient of density or velocity is present, it becomes possible to accrete angular momentum and the flows become unstable, instead of a steady state [47]. The shock cone oscillates from one side of the accretor to the other side, allowing the appearance of transient accretion disks. This instability, generally known as flip-flop oscillation [48], produces fluctuations in the accretion rate and gives rise to stochastic accretion of positive and negative angular momentum, leading to the suggestion that it is the source of the variations seen in the pulse evolution of some supergiant X-ray binaries [49, 50]. If the accretion flow is stable, the amount of angular momentum transferred to NS is negligible. The more unstable the accretion flow, the higher the transfer rate of angular momentum [48]. A secular evolution can be described by sudden jumps between states with counter-rotating quasi-Keplerian atmosphere around the magnetosphere, which imposes an inverted accretion torque on NS expressed as (25). In addition, the reversal rotation between the adiabatic atmosphere and the magnetosphere may also cause a flip-flop behavior of some clumps in the atmosphere, ejecting a part of material and leading to the loss of thermal energy, which imposes a torque, with the same form as (11) [32, 37]. Accordingly, the total spin-down torque in the regime of $r_{\mathrm{acc}} > r_{\mathrm{mag}}$ and $r_{\mathrm{cor}} > r_{\mathrm{mag}}$ is

$$T_a = -\kappa_t \frac{\mu^2}{r_{\mathrm{mag}}^3} - \dot{M} \Omega_s r_{\mathrm{mag}}^2. \quad (31)$$

Substituting it into (12), we obtain the spin evolution [43]:

$$\dot{P}_{s,\mathrm{acc}} \sim 10^{-6} I_{45}^{-1} \left[ \kappa_t \mu_{33}^{2/7} P_{s,3}^{2} \left( \frac{M}{M_\odot} \right)^{15/7} \cdot \dot{M}_{w,-6}^{6/7} v_{w,8}^{-24/7} a_{10}^{-12/7} + \mu_{33}^{8/7} P_{s,3} \left( \frac{M}{M_\odot} \right)^{-14/7} \cdot \dot{M}_{w,-6}^{3/7} v_{w,8}^{16/7} a_{10}^{8/7} \right] \mathrm{s\,s}^{-1}. \quad (32)$$

The spin-down timescale reads

$$\tau_{s,\mathrm{acc}} \sim 10^{3} I_{45} \left[ \kappa_t \mu_{33}^{2/7} P_{s,3} \left( \frac{M}{M_\odot} \right)^{15/7} \dot{M}_{w,-6}^{6/7} v_{w,8}^{-24/7} a_{10}^{-12/7} + \mu_{33}^{8/7} \left( \frac{M}{M_\odot} \right)^{-14/7} \dot{M}_{w,-6}^{3/7} v_{w,8}^{16/7} a_{10}^{8/7} \right]^{-1} \mathrm{yr}. \quad (33)$$

The corresponding X-ray luminosity in this scenario is

$$L_{x,\mathrm{acc}} \sim 10^{31} \left(\frac{M}{M_\odot}\right)^{26/7} \dot{M}_{w,-6}^{9/7} v_{w,8}^{-36/7} a_{10}^{-18/7} \mu_{33}^{-4/7} \ \mathrm{ergs\,s^{-1}}. \tag{34}$$

*2.4. Accretion from an Accretion Disk*

*2.4.1. Disk Formation.* In some systems the companion evolves and fills its Roche lobe, and a RLOF occurs. A consequence of RLOF is that the transferred material has a rather high specific angular momentum, so that it cannot be accreted directly onto the mass-capturing NS. Note that the matter must pass from the Roche lobe of the companion to that of the NS through the inner Lagrangian point. Within the Roche lobe of the NS, the dynamics of high angular momentum material will be controlled by the gravitational field of the NS alone, which would give an elliptical orbit lying in the binary plane. A continuous stream trying to follow this orbit will therefore collide with itself, resulting in dissipation of energy via shocks, and finally settles down through postshock dissipation within a few orbital periods into a ring of lowest energy for a given angular momentum, that is, a circular ring. Since the gas has little opportunity to rid itself of the angular momentum it carried, we thus expect the gas to orbit the NS in the binary plane. Such a ring will spread both inward and outward, as effectively dissipative processes, for example, collisions of gas elements, shocks, and viscous dissipation, and convert some of the energy of the bulk orbital motion into internal energy, which is partly radiated and therefore lost to the gas. The only way in which the gas can meet this drain of energy is by sinking deeper into the gravitational potential of the NS, which in turn requires it to lose angular momentum. The orbiting gas then redistributes its angular momentum on a timescale much longer than both the timescale over which it loses energy by radiative cooling and the dynamical timescale. As a result, the gas will lose as much energy as it can and spiral in towards the NS through a series of approximately Keplerian circular orbits in the orbital plane of binary, forming an accretion disk around the NS, with the gas in the disk orbiting at Keplerian angular velocity $\Omega_K(r) = (GM/r^3)^{1/2}$.

To be accreted onto the NS, the material must somehow get rid of almost all its original angular momentum. The processes that cause the energy conversion exert torques on the inspiraling material, which transport angular momentum outward through the disk. Near the outer edge of the disk some other process finally removes this angular momentum. It is likely that angular momentum is fed back into the orbital motion of a binary system by tidal interaction between the outer edge of disk and companion star. An important consequence of this angular momentum transport is that the outer edge of disk will in general be at some radius exceeding the circularization radius given by Frank et al. [9], and it is obvious that the outer disk radius cannot exceed the Roche lobe of the NS. Typically the maximum and minimum disk radii differ by a factor of 2-3. However, the disk cannot extend all the way to NS surface, because of the magnetic field, which stops the accreting plasma at a position where the pressure of field and the plasma become of the same order and leads to the truncation and an inner boundary of accretion disk.

*2.4.2. Magnetosphere.* We consider a steady and radial free-fall accretion flow, with constant accretion rate $\dot{M}$, approaching the magnetized NS. Thus the inward radial velocity and mass density near the boundary are given by

$$\begin{aligned} v_{ff} &\equiv \left(\frac{2GM}{r}\right)^{1/2} = 1.6\times 10^{9} r_8^{-1/2} \left(\frac{M}{M_\odot}\right)^{1/2} \ \mathrm{cm\,s^{-1}}, \\ \rho_{ff} &\equiv \frac{\dot{M}}{\left(v_{ff} 4\pi r^2\right)} \\ &= 4.9\times 10^{-10} r_8^{-3/2} \dot{M}_{17} \left(\frac{M}{M_\odot}\right)^{-1/2} \ \mathrm{g\,cm^{-3}}, \end{aligned} \tag{35}$$

where $r_8$ is the distance $r$ in units of $10^8$ cm.

The magnetic field of the NS begins to dominate the flow when the magnetic pressure $B^2/8\pi$ becomes comparable to the pressure of accreting matter $\sim\rho_{ff} v_{ff}^2$, which includes both ram pressure and thermal pressure. The Alfvén radius $R_A$ is then given by

$$\frac{B\left(R_A\right)^2}{8\pi} = \rho_{ff(R_A)} v_{ff}\left(R_A\right)^2, \tag{36}$$

for a dipolar field outside the NS $B(r) = \mu/r^3$. So the Alfvén radius reads

$$\begin{aligned} R_A &\sim 10^8 \dot{M}_{17}^{-2/7} \mu_{30}^{4/7} \left(\frac{M}{M_\odot}\right)^{-1/7} \ \mathrm{cm} \\ &\sim 10^8 L_{37}^{-2/7} \mu_{30}^{4/7} \left(\frac{M}{M_\odot}\right)^{-1/7} R_6^{-2/7} \ \mathrm{cm}, \end{aligned} \tag{37}$$

where $\mu_{30}$ is the magnetic moment of NS in units of $10^{30}$ G cm$^3$, and $L_{37}$ is the accretion luminosity in units of $10^{37}$ ergs s$^{-1}$.

Initially, the radial flow tends to sweep the field lines inward, due to the high conductivity of plasma. Since the plasma is fully ionized and therefore highly conducting, we would expect it to move along the field lines close enough to the NS surface. However, it is the high conductivity of accreting plasma that complicates its dynamics, since it denies the plasma ready access to the surface [51]. The plasma tends to be frozen to the magnetic flux and be swept up, building up magnetic pressure, which halts the flow and truncates the inner disk. Finally, an empty magnetic cavity is created and surrounded by the plasma that is piling up, driving the boundary inward as the plasma pressure there keeps on increasing. Such transition behavior may possibly occur at the very beginning of the accretion flow, but a steady state is quickly reached. When the magnetic pressure of the NS magnetic field can balance the ram pressure of inflows, the radial flow eventually halts, creating a restricted region inside which the magnetic field dominates the motion of the plasma. This is the magnetosphere [36], whose formation is a consequence of the transition of the plasma flow.

The boundary of the static magnetosphere is determined by the condition of static pressure balance between the inside and outside of the magnetosphere, including both plasma pressure $P$ and magnetic pressure $B^2/8\pi$; that is,

$$\left(\frac{B_t^2}{8\pi}\right)_{\text{in}} + P_{\text{in}} = \left(\frac{B_t^2}{8\pi}\right)_{\text{out}} + P_{\text{out}}, \quad (38)$$

where $B_t$ is the tangential component of magnetic field, and the subscripts "in" and "out" denote "inside" and "outside" of the magnetospheric boundary. This pressure balance relation can be used to estimate the scale size of magnetosphere in a static configuration $r^0_{\text{mag}}$, which depends on the particular physical conditions near the magnetospheric boundary [36]. However, although the supersonic flow of plasma toward the boundary is halted [8], some plasma can enter the magnetosphere via particle entry through the polar cusps, diffusion of plasma across the magnetospheric boundary, and magnetic flux reconnection [52], as well as the Rayleigh-Taylor instability at the boundary. If there is a plasma flow into the magnetosphere, its static structure is broken, and the static pressure balance (38) is no longer valid. Instead, the momentum balance and continuity of magnetic field components normal to the magnetospheric boundary imply

$$\begin{aligned} &\left(\frac{B_t^2}{8\pi}\right)_{\text{in}} + P_{\text{in}} + \left(\rho v_n^2\right)_{\text{in}} \\ &\quad = \left(\frac{B_t^2}{8\pi}\right)_{\text{out}} + P_{\text{out}} + \left(\rho v_n^2\right)_{\text{out}}, \end{aligned} \quad (39)$$

where $\rho v_n^2$ is the dynamic pressure of plasma in terms of its velocity components $v_n$ normal to the boundary. Obviously, the scale of magnetosphere is in the order of the Alfvén radius, $r_{\text{mag}} \sim R_A$.

*2.4.3. Transition Zone and Boundary Layer.* The region where the field lines thread the disk plasma forms a transition zone between the unperturbed disk and the magnetosphere, in which the disk plasma across the field lines generates currents. The currents confine the magnetic field inside a screening radius $r_s \sim (10–100)r_{\text{cor}}$, which gives the extent of transition zone. Within this region, the accreted matter is forced to corotate with the NS, by means of transporting angular momentum via stresses dominated by the magnetic field. According to the angular velocity of accreting material, the transition zone is divided into two parts (Figure 1), an outer transition zone with Keplerian angular velocity, and a boundary layer in which the plasma deviates from the Keplerian value significantly, which is separated at the radius $r_0$.

The structure of outer transition zone between $r_0$ and $r_s$ is very similar to that of a standard $\alpha$-disk [25] at the same radius, with three modifications, that is, the transport of angular momentum between disk and NS due to the magnetic stress associated with the twisted field lines, the radiative transport of energy associated with the effective viscous dissipation and resistive dissipation of currents generated by the cross-field motion of the plasma [53], and the dissipative stresses consisting of the usual effective viscous stress and the magnetic stress associated with the residual magnetospheric field. In the outer transition zone, the poloidal field is screened on a length-scale $\sim r$ by the azimuthal currents generated by the radial cross-field drift of plasma. The azimuthal pitch of the magnetic field increases from $r_0$ to a maximum at the corotation point $r_{\text{cor}}$ and begins to decrease. Thus the field lines between $r_{\text{cor}}$ and $r_s$ are swept backward and exert a spin-down torque on the NS.

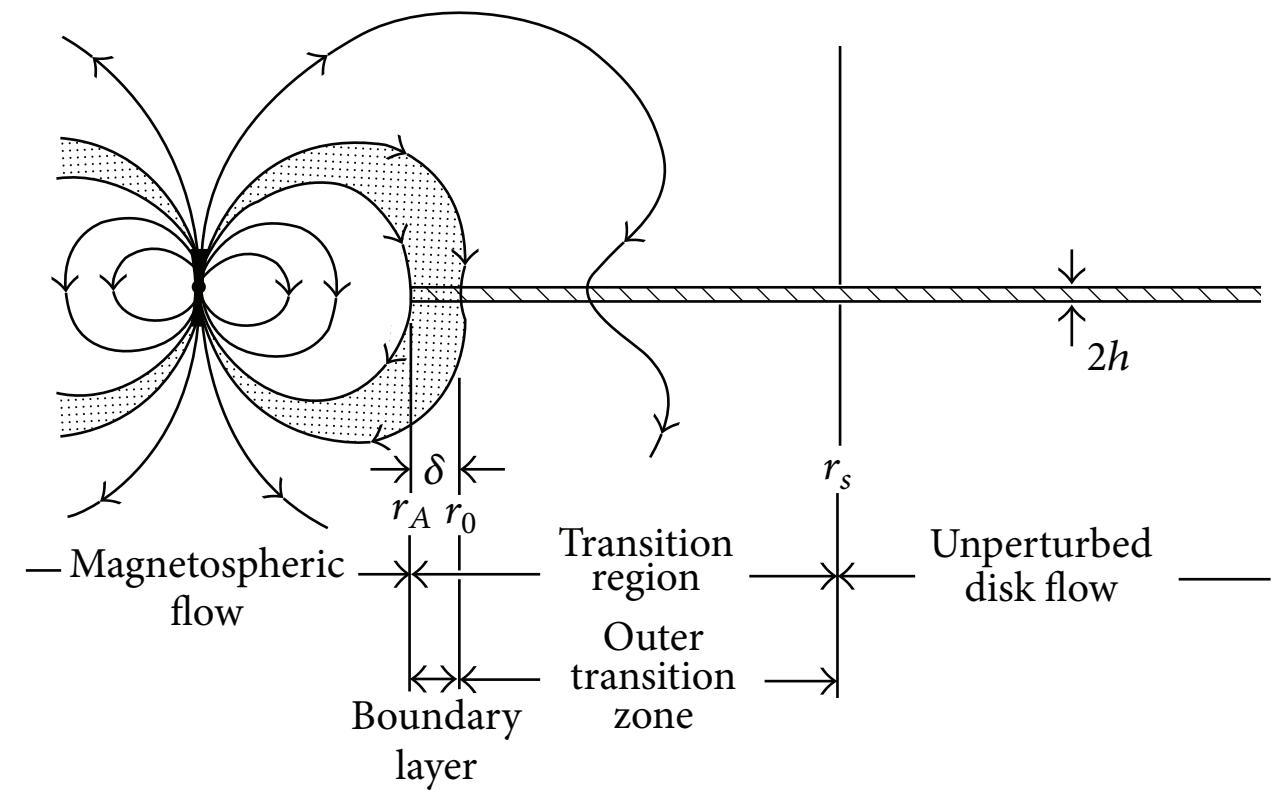

Figure 1: Schematic representation of the transition zone [41], which consists of an outer transition zone and a boundary layer.

The Keplerian motion ends at radius $r_0$, and then the angular velocity of the plasma is reduced from the Keplerian to corotate with NS, which resorts to the transportation of angular momentum through magnetic stresses [54]. Accordingly, the accreted matter is released from the disk and starts flowing toward the NS along the magnetic field lines. The transition of the accretion flow requires a region with some extent. This is the boundary layer, with thickness of $\delta \equiv r_0 - R_A \ll r_0$, in which the angular momentum is conserved, and the angular velocity of plasma continuously changes from $\Omega_K(r_0)$ to $\Omega_s$ between $r_0$ and $r_{\text{cor}}$. Because of a sub-Keplerian angular velocity, the close balance between centrifugal force and gravitational force is broken, and the radial flow attains a much higher velocity, increasing inward continuously from $r_0$ where it must equal the slow radial drift characteristic of the outer transition zone due to continuity. Since the rising magnetic pressure gradient opposes the centrifugal support, the radial velocity passes through a maximum and is reduced to zero at the inner edge of the boundary layer. The boundary layer is basically an electromagnetic layer, in the sense that the dominant stresses are magnetic stress, and the dominant dissipation is through electromagnetic processes, which obey Maxwell's equations. However, the mass flow also plays an essential role in this layer, since the cross-field radial flow generates the toroidal electric currents that screens the magnetic field of the NS.

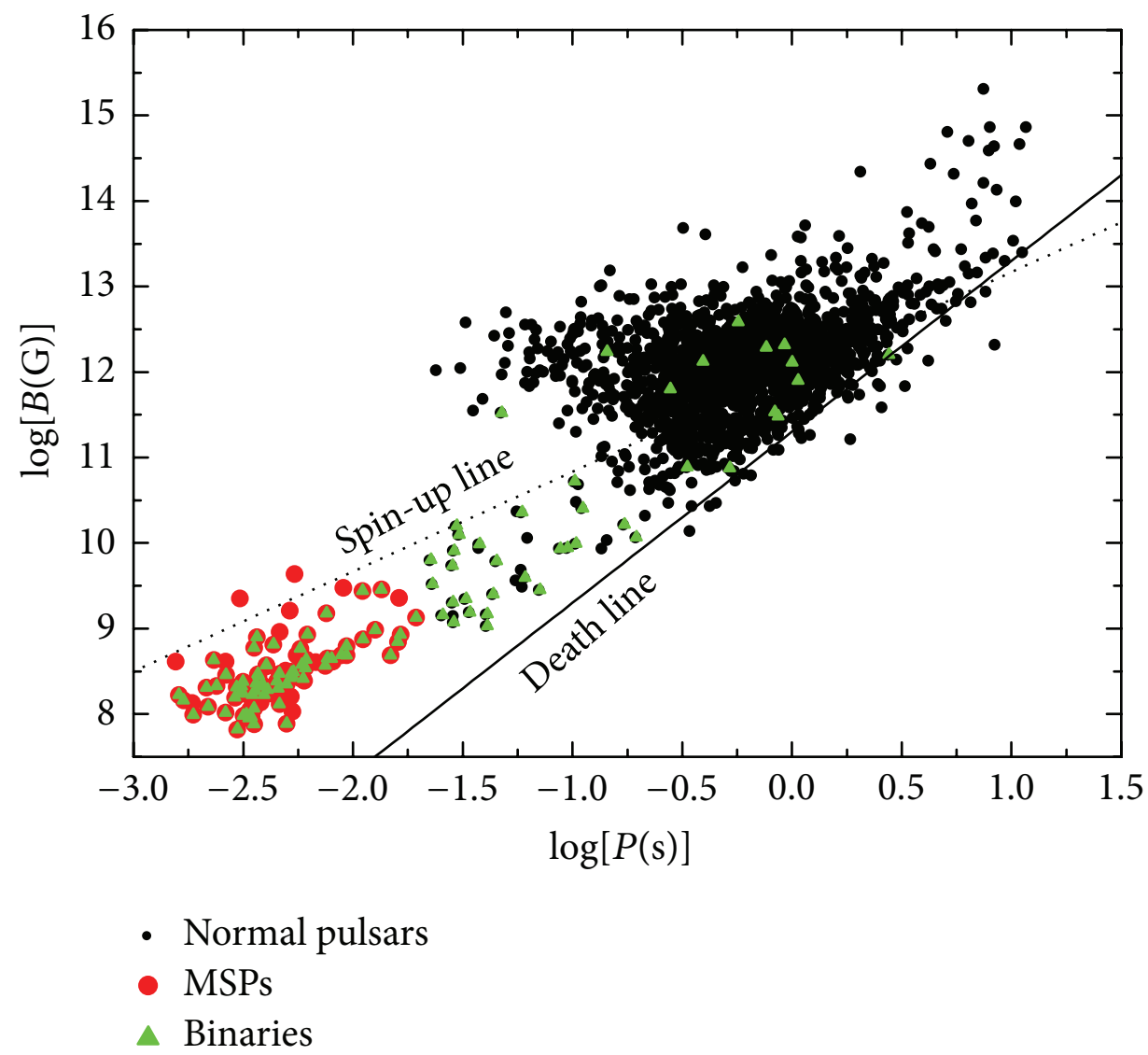

FIGURE 2: Magnetic fields and spin periods of observed pulsars (data taken from the ATNF pulsar catalogue). Black dots denote normal pulsars. Red dots represent MSPs. Green triangles are pulsars in binaries. The "spin-up line" represents the minimum spin period to which a spin-up process may proceed in an Eddington-limited accretion, while the "death line" corresponds to a polar cap voltage below which the pulsar activity is likely to switch off [17].

### 2.4.4. Accretion-Induced Field Decay and Spin-Up

*Formation of the Question.* Observationally, pulsars are usually assigned to be of two distinct kinds, whose spin periods and surface magnetic fields are distributed almost bimodally, with a dichotomy, for magnetic fields and spin periods, which are between $B \sim 10^{11-13}$ G and $B \sim 10^{8\text{-}9}$ G, respectively, and between $P_s \sim 0.1$–10 s and $P_s \sim 1$–20 ms, respectively [55]. The large population of pulsars, with high surface magnetic fields of $B \sim 10^{11-13}$ G and long spin periods of $P_s \sim$ a few seconds, represents the overwhelming majority of normal pulsars. Most of them are isolated pulsars, with only very few sources in binary systems. On the other hand, there is a smaller population of sources which display much weaker magnetic fields $B \sim 10^{8\text{-}9}$ G and short spin periods $P_s \leq 20$ ms, and most of this population is associated with binaries (Figure 2). These two typical populations are connected with a thin bridge of pulsars in binaries. The abnormal behavior of the second group, particularly the fact that these are mostly found in binaries, had led to extensive investigations concerning the formation of millisecond pulsars (MSPs). The currently widely accepted idea is the accretion-induced magnetic field decay and spin-up; that is to say the normal pulsars in binary systems have been "recycled" [56–58]. The objects formed by such a recycling process are called recycled pulsars [17, 59].

*Accretion-Induced Field Decay and Spin-Up.* The accreted matter, in the transition zone, begins to be channeled onto the polar patches of the NS by the field lines, where the compressed accreted matter causes the expansion of polar zone in two directions, downward and equatorward [60]. Therefore, the magnetic flux in the polar zone is diluted, with more and more plasma piling up onto the polar caps, which subsequently diffuse outwards over the NS surface. This process expands the area of the polar caps, until occupying the entire surface, and the magnetic flux is buried under the accreted matter. Finally, the magnetosphere is compressed to the NS surface, and no field lines drag the plasma, remaining an object with weak magnetic field. The minimum field is determined by the condition that the magnetospheric radius equals the NS radius, which is about $\sim 10^8$ G. Meanwhile, the angular momentum carried by the accreted matter increases to the spin angular momentum of the NS and spins it up, leaving a MSP with spin period of a few milliseconds. The combined field decay and spin-up process is called the recycling process of the NS.

During the recycling process, the accretion-induced magnetic field evolution can be obtained analytically for an initial field $B(t=0) = B_0$ and final magnetic field $B_f$ by [60]

$$B(t) = \frac{B_f}{\left(1 - \left[C\exp(-y) - 1\right]^2\right)^{7/4}}. \tag{40}$$

Here, we define the parameters as follows, $y = (2/7)(\Delta M/M_{\rm cr})$, the accreted mass $\Delta M = \dot{M}t$, the crust mass $M_{\rm cr} \sim 0.2M_\odot$, and $C = 1 + \sqrt{1 - x_0^2} \sim 2$ with $x_0^2 = (B_f/B_0)^{4/7}$. The bottom magnetic field $B_f$ is defined by the magnetospheric radius matching the NS radius; that is, $r_{\rm mag}(B_f) = R$. The magnetospheric radius $r_{\rm mag}$ is taken as $r_{\rm mag} = \phi R_A$ with a model dependent parameter $\phi$ of about 0.5 [9, 13, 61]. Using the relation $r_{\rm mag}(B_f) = R$, we can obtain the bottom magnetic field,

$$B_f = 1.32 \times 10^8 \left(\frac{\dot{M}}{\dot{M}_{18}}\right)^{1/2} \frac{M}{M_\odot}^{1/4} R_6^{-5/4} \phi^{-7/4}\,\text{G}, \tag{41}$$

where $\dot{M}_{18} = \dot{M}/10^{18}\,\text{g}\,\text{s}^{-1}$. For details see [60].

In the meantime, the NS is spun up by the angular momentum carried by accreted matter, according to [61]

$$\begin{aligned} -\dot{P}_s = {} & 5.8 \times 10^{-5} \left[\left(\frac{M}{M_\odot}\right)^{-3/7} R_6^{12/7} I_{45}^{-1}\right] \\ & \times B_{12}^{2/7} \left(P_s L_{37}^{3/7}\right)^2 n(\omega_s)\ \text{s}\,\text{yr}^{-1}, \end{aligned} \tag{42}$$

where $B_{12}$ is the magnetic field in units of $10^{12}$ G, and the dimensionless torque $n(\omega_s)$ is a function of the fastness parameter [36], which is introduced in order to describe the relative impotance of stellar rotation and defined by the ratio parameter of the angular velocities [36, 54],

$$\omega_s \equiv \frac{\Omega_s}{\Omega_K(R_A)} = 1.35 \left(\frac{M}{M_\odot}\right)^{-2/7} R_6^{15/7} B_{12}^{6/7} P_s^{-1} L_{37}^{-3/7}. \tag{43}$$

Therefore, the dimensionless accretion torque on NS is given by [61]

$$n(\omega_s) = 1.4 \times \left(\frac{1 - \omega_s/\omega_c}{1 - \omega_s}\right), \tag{44}$$

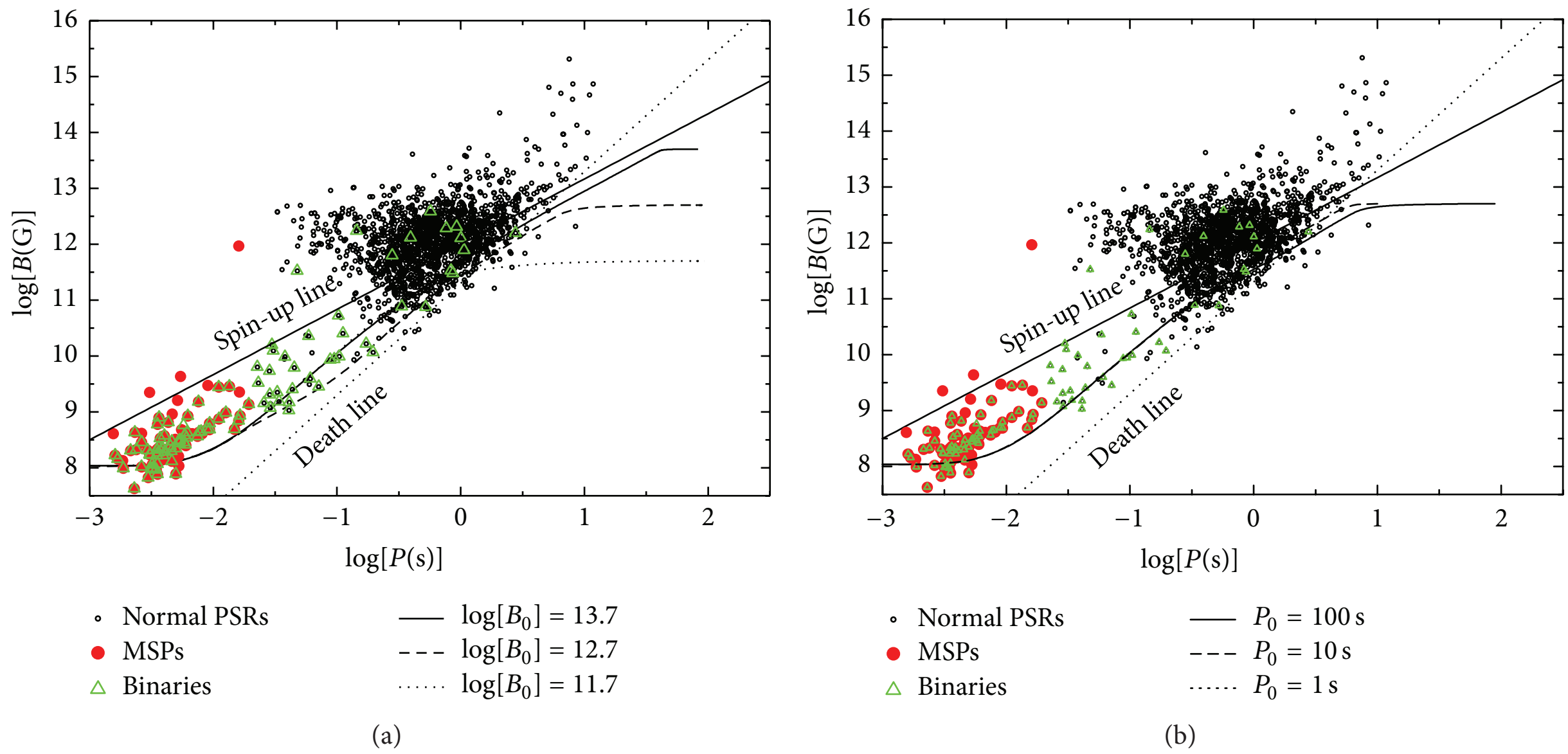

Figure 3: Magnetic field and spin period evolutionary tracks. In (a) tracks are plotted for different initial magnetic fields: $B_0 = 5 \times 10^{13}$ G (solid line), $5 \times 10^{12}$ G (dash line), and $5 \times 10^{11}$ G (dot line), at fixed initial spin period $P_0 = 100$ s and accretion rate $10^{17}$ g s$^{-1}$. In (b) with different initial periods, $P_0 = 1$ s (dotted line), 10 s (dashed line), and 100 s (solid line), at fixed initial field $B_0 = 5 \times 10^{12}$ G and accretion rate $10^{17}$ g s$^{-1}$.

where $\omega_c \sim 0.2$–1 [62–66], with an original value of 0.35 [61, 67]. For a slowly rotating star, $\omega_s \ll 1$, the torque is $n(\omega_s) \approx 1.4$, so that $T_{\rm disk} \approx 1.4T_0$. For fast rotators, $n(\omega_s)$ decreases with increasing $\omega_s$ and vanishes for the critical value $\omega_c$. For $\omega_s > \omega_c$, $n(\omega_s)$ becomes negative and increases with the increasing $\omega_s$, until it reaches a maximum value, typically, $\omega_{\rm max} \sim 0.95$.

The NS cannot be spun up to infinity because of the critical fastness $\omega_c$, at which this torque vanishes and the NS is spun up to reach the shortest spin period. Then, the rotation reaches an equilibrium spin period $P_{\rm eq}$:

$$P_{\rm eq} = 2\pi\omega_c^{-1}\Omega_K r_{\rm mag} = 1.89 B_9^{6/7}\ \text{ms}, \tag{45}$$

where $B_9$ is $B$ in units of $10^9$ G, and the second relation is obtained by taking $M = 1.4M_\odot$ and $R_6 = 1$ and an Eddington limit accretion rate [17]. This line in field-period diagram is referred to as the "spin-up line".

At the end of the accretion phase (with an accreted mass $\gtrsim 0.2M_\odot$), the magnetic field and spin period of recycled pulsars arrive at bottom values, which cluster in a range of $B = 10^{8\text{-}9}$ G and $P_s < 20$ ms [68, 69], and the NSs remain to be MSPs [70, 71]. The minimum spin period and bottom field are independent of the initial values of magnetic field and spin period; see Figure 3 [72]. The bottom fields mildly vary with the accretion rates and the accreted mass (Figure 4), while the minimum period is insensitive to them (Figure 5) [72, 73].

The direct evidence for this recycled idea has been found in LMXBs with an accreting millisecond X-ray pulsar, for example, SAX J 1808.4-3658 [74], and in observing the link between LMXBs and millisecond radio pulsars in the form of the transition from an X-ray binary to a radio pulsar PSR J 1023+0038 [75]. The NSs in LMXBs are the evolutionary precursors to "recycled" MSPs [57]. It is evident that X-ray pulsars and recycled pulsars are correlated with both the duration of accretion phase and the total amount of accreted matter [57, 76]. When a small amount of mass ($\lesssim 0.001M_\odot$) is transferred, the spin period mildly changes, which may yield a HMXB with a spin period of some seconds, such as Her X-1 [77] and Vela X-1 [78]. If the NS accretes a small quantity of mass from its companion, for example, $\sim 0.001M_\odot$–$0.01M_\odot$, a recycled pulsar with a mildly weak field ($B \sim 10^{10}$ G) and short spin period ($P_s \sim 50$ ms) will be formed [79], like PSR 1913+16 and PSR J 0737-3039 [80]. After accreting sufficient mass ($\Delta M \gtrsim 0.1M_\odot$), the lower magnetic field and shorter spin period ($P_s \lesssim 20$ ms) of a MSP form, for example, SAX J 1808.4-3658 [74] and PSR J 1748-2446 [81].

*2.4.5. Rotation Effects and Quasi-Quantized Disk.* The orbital motion of an accreting flow controlled by the potential of a rotating NS is different from that in a flat space-time, because of the rotation effects and the strong gravity of the NS. Firstly, stellar oblateness arises from the rotation and thus a quadruple term in the gravitational potential appears [82]. Secondly, a rotating massive object will impose a rotational frame-dragging effect on the local inertial frame, which is known as gravitoelectromagnetism [83–86].

*Gravitoelectromagnetism.* Gravitoelectromagnetism (GEM) is based on an analogy between Newton's law of gravitation and Coulomb's law of electricity. The Newtonian solution of the gravitational field can be alternatively interpreted as a gravitoelectric field. It is well known that a magnetic field is produced by the motion of electric charges, that is, electric current. Accordingly, the rotating mass current would give rise to a gravitomagnetic field. In the framework of the general theory of relativity, a non-Newtonian massive mass-charge current can produce a gravitoelectromagnetic field [87–89].

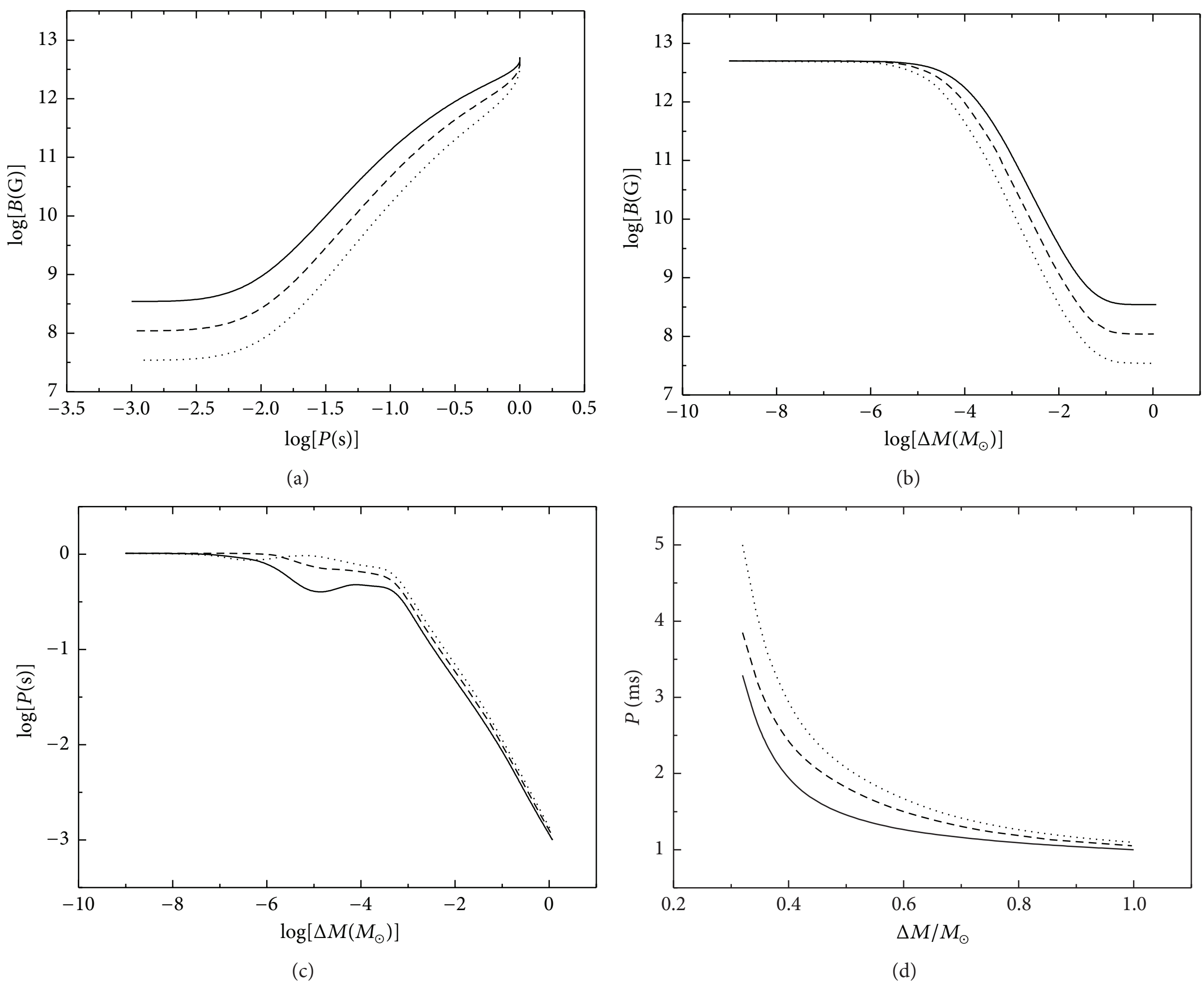

FIGURE 4: $B$ and $P$ evolution in the recycling process. (a) shows the joint evolution of $B$ and $P$. (b) and (c) are their evolution as a function of accreted mass $\Delta M$. The solid, dashed, and dotted lines are plotted for an accretion rate of $10^{18}\,\mathrm{g\,s^{-1}}$, $10^{17}\,\mathrm{g\,s^{-1}}$, and $10^{16}\,\mathrm{g\,s^{-1}}$, respectively. Initial values $B_0 = 5\times 10^{12}$ G and $P_0 = 1$ s were taken. (d) is a zoom-in view of (c) in a linear scale for spin periods shorter than 5 ms.

In the linear approximation of the gravitational field, the Minkowski metric $\eta_{\mu\nu}$ is perturbed due to the presence of a gravitating source with a perturbative term $h_{\mu\nu}$, and thus the background reads

$$g_{\mu\nu} = \eta_{\mu\nu} + h_{\mu\nu}, \quad \left|h_{\mu\nu}\right| \ll 1. \tag{46}$$

We define the trace-reversed amplitude $\overline{h}_{\mu\nu} = h_{\mu\nu} - (1/2)\eta_{\mu\nu}h$, where $h = \eta^{\mu\nu}h_{\mu\nu} = h^{\alpha}_{\alpha}$ is the trace of $h_{\mu\nu}$. Then, expanding the Einstein's field equations $G_{\mu\nu} = 8\pi G T_{\mu\nu}$ in powers of $\overline{h}_{\mu\nu}$ and keeping only the linear order terms, we obtain the field equations (in this part, we set the light speed $c = 1$) [82]:

$$\Box \overline{h}_{\mu\nu} = -16\pi G T_{\mu\nu}, \tag{47}$$

where the Lorentz gauge condition $\overline{h}^{\mu\nu}_{,\nu} = 0$ is imposed. In analogy with Maxwell's field equations $\Box A^{\nu} = 4\pi j^{\nu}$, we can find that the role of the electromagnetic vector potential $A^{\nu}$ is played by the tensor potential $\overline{h}_{\mu\nu}$, while the role of the 4-current $j^{\nu}$ is played by the stress-energy tensor $T_{\mu\nu}$. Therefore, the solution of (47) in terms of the retarded potential can be written as [90, 91]

$$\overline{h}_{\mu\nu} = 4G \int \frac{T_{\mu\nu}\left(q_0 - \left|q_i - q_i'\right|, q_i'\right)}{\left|q_i - q_i'\right|} d^3 q_i', \tag{48}$$

with coordinates $q_{\mu} = (q_0, q_i)$. (Note that we choose Greek subscripts and indices (i.e., $\mu$, $\nu$, $\alpha$, $\beta$) to describe the 4-dimension space-time components $(0, 1, 2, 3)$ for test particles, while the subscripts and indices $i$, $j$, $k$ denote the 3-dimension space coordinates $x$, $y$, $z$ (or $r$, $\theta$, $\phi$).)

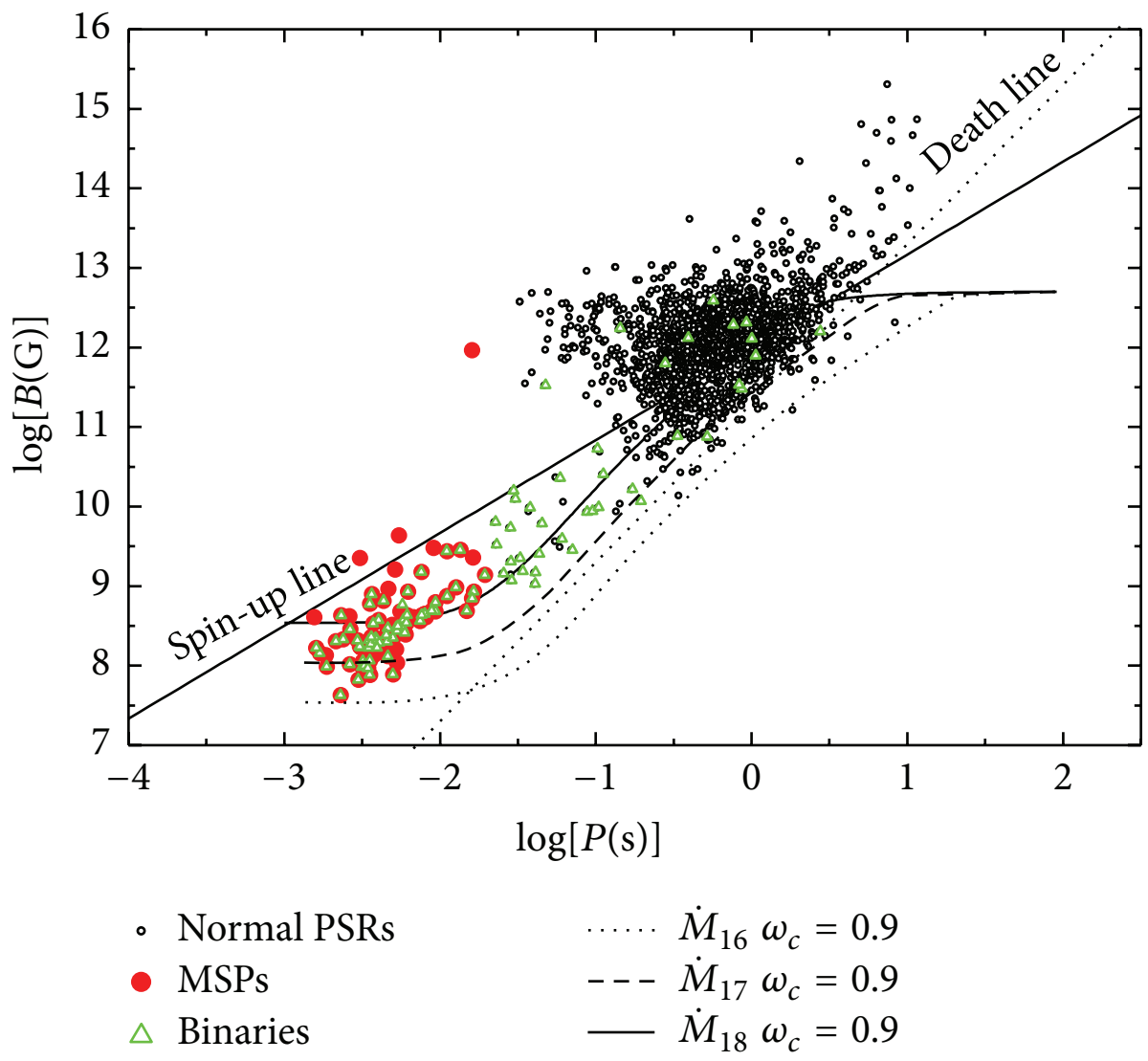

Figure 5: Magnetic field and spin period evolutionary tracks for different accretion rates: $\dot{M} = 10^{18}\,\mathrm{g\,s^{-1}}$ (solid line), $\dot{M} = 10^{17}\,\mathrm{g\,s^{-1}}$ (dashed line), and $\dot{M} = 10^{16}\,\mathrm{g\,s^{-1}}$ (dotted line), at fixed initial field $B_0 = 5\times10^{12}$ G and period $P_0 = 100$ s.

Neglecting all terms of high order and smaller terms that include the tensor potential $\overline{h}_{ij}(q_0, q_i)$ [82], we can write down the tensor potential $\overline{h}_{\mu\nu}$ [91]:

$$\overline{h}_{00} = 4\Phi, \qquad \overline{h}_{0i} = -2A_i. \tag{49}$$

Here, $\Phi$ and $A_i$ are the Newtonian or gravitoelectric potential and the gravitomagnetic vector potential, respectively. They can be expressed as

$$\Phi(q_0, q_i) = -\frac{GM}{r}, \qquad A_i(q_0, q_i) = G\frac{J_j q_k}{r^3}\varepsilon_i^{jk}, \tag{50}$$

where $A_i$ is in terms of the angular momentum $J_i$ of the NS. Accordingly, the Lorentz gauge condition reduces to

$$\frac{\partial\Phi}{\partial q_0} + \frac{1}{2}\nabla_i A_j \delta_{ij} = 0. \tag{51}$$

Then we can define the gravitoelectromagnetic field, that is, the gravitoelectric field $g_i$ and the gravitomagnetic field $\mathfrak{R}_i$, as [90]

$$g_i = -\nabla_i\Phi - \frac{\partial}{\partial q_0}\left(\frac{1}{2}A_j\right)\delta_{ij}, \qquad \mathfrak{R}_i = \nabla_j A_k \varepsilon_i^{jk}. \tag{52}$$

In the approximation of weak field [82, 92] and slow rotation of the NS, $g_i$ contains mostly first-order corrections to flat space-time and denotes the Newtonian gravitational acceleration, whereas $\mathfrak{R}_i$ contains second-order corrections and is related to the rotation of the NS.

Using (47), (51), and (52) and analogy with Maxwell's equations, we get the gravitoelectromagnetic field equations [90]:

$$\nabla_i g_j \delta_{ij} = -4\pi G\rho_g, \qquad \nabla_j g_k = -\frac{\partial}{\partial q_0}\left(\frac{1}{2}\mathfrak{R}_i\right), \qquad \nabla_i \mathfrak{R}_j \delta_{ij} = 0, \qquad \nabla_j \mathfrak{R}_k = \frac{\partial g_i}{\partial q_0} - 4\pi G j_{gi}. \tag{53}$$

Here, $\rho_g$ is the mass density, and $j_g = \rho_g\times$ (velocity of the mass flow generating the gravitomagnetic field) denotes the mass current density or mass flux. These equations include the conservation law for mass current $\partial\rho/\partial q_0 + \nabla_i j_{gj}\delta_{ij} = 0$, as they should be.

In analogy with the Lorentz force in an electromagnetic field, the motion of a test particle $m$ in a gravitoelectromagnetic field is subject to a gravitoelectromagnetic force

$$F_{\mathrm{GEM}i} = Q_g g_i + Q_{\mathfrak{R}}\dot{q}_j \mathfrak{R}_k \varepsilon_i^{jk}, \tag{54}$$

where $Q_g = -m$ and $Q_{\mathfrak{R}} = -2m$ are the gravitoelectric charge and gravitomagnetic charge of the test particle [90], respectively. Accordingly, the Larmor quantities in the gravitoelectromagnetic field would be

$$\vec{a}_L = -\frac{Q_g\vec{g}}{m}, \qquad \vec{\omega}_L = \frac{Q_{\mathfrak{R}}\vec{\mathfrak{R}}}{2m}. \tag{55}$$

Therefore, the test particle has the translational acceleration $\vec{a}_L = \vec{g}$ and the rotational frequency $\omega_L = |\vec{\mathfrak{R}}|$.

For a rotating NS with angular velocity $\Omega_s$, the gravitational Larmor frequency of test particles can be written as [93]

$$\omega_L = \left|\vec{\mathfrak{R}}\right| = \frac{4}{5}\frac{GMR^2}{r^3}\Omega_s, \tag{56}$$

which causes a split of the orbital motion of accreted particles and changes the circular orbital motion on the binary plane in the vertical direction of disk.

*Axisymmetrically Rotating Stars.* The stationary and axisymmetric space-time metric arising from a rotating object, in the polar coordinates, has a form of [94]

$$ds^2 = e^{2\nu(r,\theta)}dt^2 - e^{2\lambda(r,\theta)}dr^2 - e^{2\mu(r,\theta)}\left[r^2 d\theta^2 + r^2\sin^2\theta\,(d\phi - \omega(r,\theta)\,dt)^2\right]. \tag{57}$$

As a new feature of an axisymmetric structure, the nondiagonal elements appear:

$$g_{t\phi} = g_{\phi t} = r^2 \sin^2\theta e^{2\mu(r,\theta)} \omega(r,\theta). \quad (58)$$

For a test particle at a great distance from NS in its equatorial plane ($\theta = \pi/2$) falling inwards from rest, the $\phi$ coordinate at $\theta = \pi/2$ obeys the equation [95]:

$$\frac{d\phi}{dt} = \frac{g^{\phi t}}{g^{tt}} = \frac{-g_{\phi t}}{g_{\phi\phi}} = \frac{-r^2 e^{2\mu(r,\theta)} \omega(r)}{-r^2 e^{2\mu(r,\theta)}} = \omega(r). \quad (59)$$

Therefore, $\omega(r)$ is referred to as the angular velocity of the local inertial frame, which can be expressed in terms of the Keplerian velocity at radius $r$ [96]:

$$\omega(r) = \frac{2}{5}\sqrt{\frac{GM}{r^3}} R^2 \Omega_s^2, \quad (60)$$

Consequently, the accreted matter experiences an increasing drag in the direction of the rotation of NS and possesses a corresponding angular velocity $\omega(r)$.

*Closure Orbits and the Quasi-Quantized Disk*. The accretion flow in a turbulent Keplerian disk will flow helical trajectories, that is, open circular orbits at each radius, and is endowed with two frequencies, that is, angular frequency in the direction of rotation and gravitational Larmor frequency in vertical direction of the disk. The former deviates the orbital motion from a circular orbit, and the latter leads to some vertical oscillation. If and only if the gravitational Larmor frequency and angular velocity of accreted plasma satisfy [97]

$$l\omega(r) = n\omega_L, \quad (61)$$

where $n$ and $l$ are integers with $n, l \geq 1$, the vertical split along disk and deformation in the direction of rotation of NS can be harmonious [98], that is, the vertical oscillation and deformed orbital circular motion are syntonic. As a result, the orbital motion of accreted matter on separated and deformed open circular orbits returns to a closed circular motion. The radii of closure circular orbits are give by

$$r^3 = \frac{4n^2}{l^2}\frac{GM}{\Omega_s^2}. \quad (62)$$

We call such a disk structure, that is, tenuous spiraling-in gas with the appearance of a band of closure circular orbits, as a quasi-quantized disk. Moving on the closure circular orbits, the accreted material is in a stable state, with the first derivative of angular momentum being larger than or equal to zero, which correspond to a minimum of the effective potential. With a slight perturbation, the test particle will oscillate around the minimum, manifesting as drift of the orbital frequency. If the perturbation is strong enough to transfer sufficient angular momentum outwardly and drive the particles to leave this state, the material will continue to follow the original helical track and spirals in. In a turbulent and viscous accretion disk, dissipative processes, for example, viscosity, collisions of elements, and shocks, are responsible for the perturbation.

## 3. Final Remarks

In NS X-ray binary systems, accretion is the only viable energy source that powers the X-ray emission. Accretion onto a NS is fed by mass transfer from the optical companion star to the NS, via either a stellar wind or RLOF. Stellar wind accretion always occurs in NS/HMXBs, with high-mass OB supergiant companions or Be stars with radiation-driven stellar winds. The neutron stars here are observed as X-ray pulsars, while disk accretion dominates the mass-transfer mechanism in NS/LMXBs, which occurs when the outer layer of the companion flows into the Roche lobe of the NS along the inner Lagrangian point. However, the transferred material cannot be accreted all the way onto NS surface, due to the strong magnetic field. At a preferred radius where the magnetic pressure can balance the ram pressure of the infalling flows, the accreting plasma will halt and interact with the magnetic field, building up a magnetosphere, which exerts an accretion torque and results in the spin period evolution of NS. In NS/HMXBs, because of the disordered stellar wind, some instabilities and turbulence lead to fluctuations in the mass accretion rate and give rise to a stochastic accretion with positive and negative angular momentum, which alternatively contributes to phases of spin-up and spin-down. For disk-fed LMXBs, the NSs usually are spun up by the angular momentum carried by the accreted matter, during which the magnetic field is buried by the accumulated plasma. As a result, the NS is spun up to a few milliseconds and the magnetic field decays to $10^{8\text{-}9}$ G, remaining a MSP. In several systems, the torque reversal [99–102] and state changes [103, 104] were observed and investigated because of sudden dynamical changes triggered by a gradual variation of mass accretion rates [105].

In the inner region of an accretion disk, the accreting plasma is endowed with two additional frequencies due to frame-dragging effects arising from the rotation of the NS, which contribute to a band of closed circular orbits at certain radii. Such a disk structure is expected to be the quasi-quantized structure. The orbiting material, moving on these closure circular orbits, is in a stable state. With a slight perturbation, the test particle will oscillate around the minimum, manifesting as radial drift of the orbital frequency. If the perturbation is strong enough to transfer sufficient angular momentum outwardly and drive the particles to leave this state, the material will continue to follow the helical track and spirals in.

## Competing Interests

The author declares that he has no competing interests.

## Acknowledgments

This work is supported by the Fundamental Research Funds for the Central Universities (Grant no. 161gpy49) at Sun Yat-Sen University.